\documentclass[pra,showpacs,twocolumn]{revtex4-1}%
\usepackage{amsfonts}
\usepackage{amsmath}
\usepackage{amssymb,amscd}
\usepackage{psfrag}
\usepackage{graphicx}
\usepackage{braket}
\usepackage[dvips]{epsfig}
\usepackage{epsfig}
\usepackage{subfigure}
\usepackage{color}
\usepackage{amssymb}%
\providecommand{\U}[1]{\protect\rule{.1in}{.1in}}
\begin{document}

\title{Nonlinear phase estimation enhanced by an actively correlated Mach-Zehnder interferometer}
\author{Gao-Feng Jiao$^{1}$}
\author{Keye Zhang$^{1}$}
\author{L. Q. Chen$^{1}$}
\email{lqchen@phy.ecnu.edu.cn}
\author{Weiping Zhang$^{2,3}$}
\author{Chun-Hua Yuan$^{1,3}$}
\email{chyuan@phy.ecnu.edu.cn}

\address{$^1$State Key Laboratory of Precision Spectroscopy, Quantum Institute for Light and Atoms, Department of Physics, East China Normal University, Shanghai 200062, China}
\address{$^2$School of Physics and Astronomy, and Tsung-Dao Lee Institute, Shanghai Jiao Tong University, Shanghai 200240, China}

\address{$^{3}$Collaborative Innovation Center of Extreme Optics, Shanxi University, Taiyuan, Shanxi 030006, China}

\begin{abstract}
A nonlinear phase shift is introduced to a Mach-Zehnder interferometer (MZI),
and we present a scheme for enhancing the phase sensitivity. In our scheme, one input port of a
standard MZI is injected with a coherent state and the other input port is injected with one mode of
a two-mode squeezed-vacuum state. The final interference output of the MZI is detected with the method of active correlation output readout. Based on the optimal splitting ratio of beam splitters, the phase sensitivity can beat the standard
quantum limit and approach the quantum Cram\'{e}r-Rao bound. The effects of
photon loss on phase sensitivity are discussed. Our scheme can also provide
some estimates for units of $\chi^{(3)}$, due to the relation between the
nonlinear phase shift and the susceptibility $\chi^{(3)}$ of the Kerr medium.
\end{abstract}
\date{\today }
\maketitle

\section{Introduction}

As fundamental devices, interferometers play a very important role in the
field of precision measurement \cite{Fixler J B07,Peters A01,Abbott B
P16,Abbott B P17,Abadie J11}. To date, a number of interferometer
configurations have been proposed for precision measurement. The most widely
studied configuration is the Mach-Zehnder interferometer (MZI). One of the
most important performances of an interferometer is the sensitivity of the
measurement. However, the sensitivity of interferometer measurements is
limited by the shot-noise limit, or the standard quantum limit
(SQL)\ with respect to classic resources. Therefore, researchers are generally
concerned with how to improve the sensitivity of interferometers as much as possible.

Considering the vacuum fluctuation entering from the unused input port, Caves
\cite{Caves81} proposed the squeezed-state technique, which is used to
overcome the SQL. Soon after, various quantum resources
\cite{Xiao87,NOON,Steinlechner S13} were used to improve the measurement
precision and offer the possibility of reaching the Heisenberg limit.
Yurke \cite{Yurke86} \textit{et al.} theoretically introduced the SU(1,1)
interferometer using a nonlinear beam splitter (NBS) in place of a linear
beam splitter (BS) for wave splitting and recombination, where the NBS was
provided by the optical parameter amplifier process or the four-wave mixing
process. Due to the quantum destructive interference in the SU(1,1)
interferometer, the noise accompanied by the amplification of the signal can
revert to the level of input. Benefiting from that, the signal-to-noise ratio
improves. Because it can be used to improve measurement accuracy, this type
of interferometers has received extensive attention both experimentally
\cite{Jing11,Hudelist14,Chen15,Qiu16,Linnemann16,Lemieux16,Manceau17,Anderson17,Gupta17,Du18}
and theoretically
\cite{Plick10,Ou12,Marino12,Li14,Gabbrielli15,Chen16,Sparaciari16,Li16,Gong17,Giese17,Hu18}%
. However, there exists the disadvantage that the number of phase-sensing
photons is too small, which imposes constraints on the enhancement
measurement. More recently, the pumped-up SU(1,1) interferometer \cite{Pump
up} was proposed to mitigate the situation, which may be implemented in spinor
Bose-Einstein condensates or hybrid atom-light systems. The seeded SU(1,1)
interferometer with a coherent state boost is practical to implement
experimentally \cite{Liu18}. Seed light injection can indeed increase the
number of photons in the interferometer. However, due to the limitation of the
four-wave mixing process, the number of photons increased is limited because
of the additional noise \cite{Boyd1,Boyd2,Boyd3}. This uncorrelated noise
grows as the intensity of the seed light increases, which imposes constraints
on the phase-sensing light power. Due to the progress made in theory and experiment on the SU(1,1) interferometers, Caves reframed the SU(1,1) interferometry and brought a wide variety of SU(1,1)-based measurement techniques together \cite{Caves20}.

Both the employment of exotic quantum states and the improvement on hardware
structures can enhance the phase sensitivity of an interferometer. Also,
nonlinear transformation can provide better sensitivity than linear
transformation for the phase-encoding process. The nonlinear transformation of
phase shift $\hat{U}(\phi)=e^{i\phi(\hat{a}^{\dagger}\hat{a})^{k}}$ ($k\geq2$)
can be implemented by propagating in nonlinear crystals. For example, the Kerr
effect provides the case $\hat{U}(\phi)=e^{i\phi(\hat{a}^{\dagger}\hat{a}%
)^{2}}$. Beltr\'{a}n and Luis proposed that encoding the signal via nonlinear
transformation can improve the precision and robustness of the detection scheme
\cite{N1}. Boixo \textit{et al.} showed that it is possible to achieve
measurement precision that scales better than $N^{-1}$ by using the dynamics
generated by nonlinear Hamiltonians \cite{N2}. Napolitano \textit{et al.}
experimentally realized a system designed to achieve metrological sensitivity
beyond $N^{-1}$, using nonlinear interactions among particles \cite{N4}.
Because this new perspective would greatly improve the measurement precision,
recently, nonlinear phase estimation has drawn considerable interest. Joo
\textit{et al.} investigated the phase enhancement of quantum states subject
to nonlinear phase shifts \cite{N6}. Berrada studied the phase estimation of
entangled SU(1,1) coherent states resulting from a generalized nonlinearity of
the phase shifts \cite{N7}. Cheng analyzed the quantum uncertainty bounds for
simultaneously detecting linear and nonlinear phase shifts \cite{N8}.
Zhang \textit{et al.} investigated second-order nonlinear phase estimation
using a coherent state and parity measurement \cite{N9}.

In this paper, we also introduce a nonlinear phase shift in an MZI and propose
a scheme to improve the sensitivity based on active correlation output
readout. The phase sensitivity is studied theoretically and it can beat the
SQL and approach the quantum Cram\'{e}r-Rao bound (QCRB). Our scheme can also be thought of as inserting an
MZI into one of the arms of the SU(1,1) interferometer \cite{Du20}. Compared
to a conventional SU(1,1) interferometer, the intrinsic problem of a small
number of phase-sensing photons is solved.

Our paper is organized as follows. In Sec. II, we describe the input-output
relation of the MZI with a coherent state in one input port and one mode of a
two-mode squeezed-vacuum state in the other input port and with active
correlation output readout. In Sec. III, the phase sensitivity is studied with
the method of homodyne detection and the results obtained are compared with the
QCRB. The optimal splitting ratio of beam splitters for nonlinear phase
estimation is given. In Sec. IV, the phase sensitivity in the presence of loss
is presented and discussed. Finally, we conclude with a summary of our results.

\begin{figure}[tb]
\centering{\includegraphics[scale=0.3,angle=0]{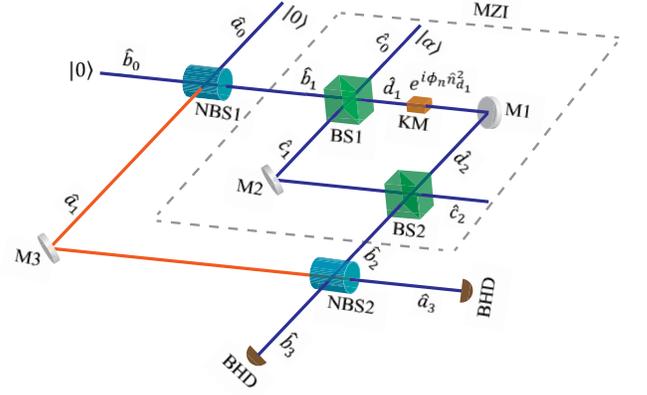}} \caption{In the
dashed box, a standard MZI is input with a coherent state in one input port
and one mode of a two-mode squeezed-vacuum state generated by NBS1 in the
other input port and is detected with active correlation readout through NBS2.
A Kerr-type medium is embedded into one path of the MZI to use as a phase
shifter, and the phase transform is $\hat{U}(\phi)=e^{i\phi_{l}\hat{n}_{d_{1}%
}+i\phi_{n}\hat{n}_{d_{1}}^{2}}$, where $\phi_{l}$ and $\phi_{n}$ represent
the linear and nonlinear phase shift, respectively. $\hat{a}_{i}$, $\hat
{b}_{i}$, $\hat{c}_{i}$, and $\hat{d}_{i}$ ($i=0,1,2,3$) denote light beams in
the different processes. $\mathrm{M}$, mirrors; $\mathrm{BS}$, linear beam
splitters; $\mathrm{NBS}$, nonlinear beam splitters; $\mathrm{BHD}$, balanced
homodyne detection.}%
\label{fig1}%
\end{figure}

\section{Model}

As shown in Fig.~\ref{fig1}, the MZI is in the dashed box, where one input
port is injected with a strong field, and another input port is sent with one
mode of a two-mode entanglement state generated by NBS1. A Kerr-type medium is
embedded into one path of the MZI to use as a phase shifter and introduce the
nonlinear phase shift. One of the output fields of the MZI and the other mode
of the two-mode squeezed state are combined to realize active correlation
output readout.

The four modes in the scheme are described by the annihilation operators
$\hat{a}_{i}$, $\hat{b}_{i}$, $\hat{c}_{i}$, and $\hat{d}_{i}$ ($i=0,1,2,3$).
The field $\hat{d}_{1}$ passing through the phase shifter will experience both
linear and nonlinear phase shifts. The corresponding phase transformation is
written as%
\begin{equation}
\hat{U}(\phi_{l},\phi_{n})=\exp(i\phi_{l}\hat{n}_{d_{1}}+i\phi_{n}\hat
{n}_{d_{1}}^{2}),
\end{equation}
where $\phi_{l}$ and $\phi_{n}$ represent the linear and nonlinear phase
shift, respectively. $\hat{n}_{d_{1}}$ denotes the photon number operator.

In the Heisenberg picture, after the phase shift the field $\hat{d}_{1}$ is
transformed as%
\begin{equation}
\hat{d}_{2}=\exp[i\phi_{l}+i\phi_{n}(2\hat{n}_{d_{1}}+1)]\hat{d}_{1},
\end{equation}
\ The two-port input-output relation of the MZI is given by%
\begin{equation}
\hat{b}_{2}=\mathcal{M}_{1}\hat{b}_{1}+\mathcal{M}_{0}\hat{c}_{0},\text{ }%
\hat{c}_{2}=\mathcal{M}_{2}\hat{c}_{0}+\mathcal{M}_{0}\hat{b}_{1},
\end{equation}
where
\begin{align}
&  \mathcal{M}_{0}=\sqrt{TR}[e^{i\phi_{l}+i\phi_{n}(2\hat{n}_{d_{1}}%
+1)}-1],\mathcal{M}_{1}=R\nonumber\\
&  +Te^{i\phi_{l}+i\phi_{n}(2\hat{n}_{d_{1}}+1)},\mathcal{M}_{2}=Re^{i\phi
_{l}+i\phi_{n}(2\hat{n}_{d_{1}}+1)}+T.
\end{align}
$R$ and $T$ are the reflectivity and transmissivity of the two BSs, respectively.

Considering one mode of a two-mode entangled state input to the MZI, the
scheme is transformed into a three-port input-output interferometer and their
relation is described by%
\begin{align}
\hat{a}_{3}  &  =\hat{a}_{0}\mathcal{A}+\hat{b}_{0}^{\dagger}\mathcal{B}%
+\hat{c}_{0}^{\dagger}\mathcal{C},\text{ }\hat{b}_{3}=\mathcal{D}\hat{a}%
_{0}^{\dagger}+\mathcal{E}\hat{b}_{0}+\mathcal{F}\hat{c}_{0},\\
\hat{c}_{2}  &  =\mathcal{M}_{2}\hat{c}_{0}+\mathcal{H}\hat{b}_{0}%
+\mathcal{I}\hat{a}_{0}^{\dagger},
\end{align}
where
\begin{align}
\mathcal{A}  &  =G_{2}G_{1}+g_{2}g_{1}e^{i\left(  \theta_{2}-\theta
_{1}\right)  }\mathcal{M}_{1}^{\ast},\text{ }\mathcal{B}=G_{2}g_{1}%
e^{i\theta_{1}}\nonumber\\
&  +G_{1}g_{2}e^{i\theta_{2}}\mathcal{M}_{1}^{\ast},\text{ }\mathcal{C}%
=g_{2}e^{i\theta_{2}}\mathcal{M}_{0}^{\ast},\mathcal{H=}G_{1}\mathcal{M}%
_{0},\nonumber\\
\text{ }\mathcal{D}  &  =G_{1}g_{2}e^{i\theta_{2}}+G_{2}g_{1}e^{i\theta_{1}%
}\mathcal{M}_{1},\text{ }\mathcal{E}=g_{2}g_{1}e^{i\left(  \theta_{2}%
-\theta_{1}\right)  }\nonumber\\
&  +G_{2}G_{1}\mathcal{M}_{1},\text{ }\mathcal{F}=G_{2}\mathcal{M}_{0},\text{
}\mathcal{I=}g_{1}e^{i\theta_{1}}\mathcal{M}_{0}. \label{eq3}%
\end{align}
Here, $G_{1}$ and $G_{2}$ are the gain factors of NBS$1$ and NBS$2$,
respectively, for wave splitting and recombination with $G_{i}^{2}-g_{i}%
^{2}=1$ ($i=1,2$). $\theta_{1}$ and $\theta_{2}$ describe the phase shift of the
NBS for wave splitting and recombination, respectively.

\begin{figure}[ptb]
\centering{\includegraphics[scale=0.4,angle=0]{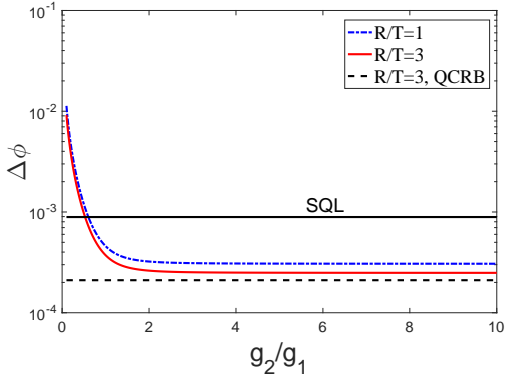}} \caption{(Color
online) The phase sensitivity as a function of $g_{2}/g_{1}$ with different
$R/T$ where $|\alpha|=10$ and $g_{1}=2$.}%
\label{fig2}%
\end{figure}

\section{Estimation of nonlinear phase shift}

The Kerr effect in media is usually interpreted as modulation of the
refractive index due to the application of a strong drive field. For many
materials, the refractive index in the presence of a drive field can be
described by \cite{Boyd}%
\begin{equation}
n=n_{0}+n_{2}\langle\hat{I}\rangle,
\end{equation}
where $n_{0}$ is the refractive index, $n_{2}$ is a nonlinear coefficient
which is proportional to the third-order susceptibility $\chi^{(3)}$ and can
be written as $n_{2}=3\chi^{(3)}/4n_{0}^{2}\epsilon_{0}c$, and $\langle\hat
{I}\rangle$ is the intensity of the drive field. In our model, the expression of
nonlinear phase shift is $\phi_{n}=3\chi^{(3)}\langle\hat{I}_{d_{1}}\rangle
k_{d1}L/4n_{0}^{2}\epsilon_{0}c$ with $L$ Kerr media width.

Here, we mainly study the nonlinear phase shift $\phi_{n}$ caused by nonlinear
susceptibility $\chi^{(3)}$ changes, since the linear phase shift $\phi_{l}$
is constant. The conceptual understanding of nonlinear optics is often based
on the use of the nonlinear susceptibility $\chi^{(3)}$ \cite{Boyd14}. Our
scheme can give the magnitude of nonlinear susceptibility through the
measurement of nonlinear phase shift. The accuracy of nonlinear susceptibility
$\chi^{(3)}$ of the Kerr medium is given by phase sensitivity $\Delta\phi_{n}$
measurement, that is%
\begin{equation}
\Delta\chi^{(3)}=\frac{4n_{0}^{2}\epsilon_{0}c}{3\langle\hat{I}_{d_{1}}\rangle
k_{d1}L}\Delta\phi_{n}.
\end{equation}
High sensitivity phase measurement will result in high-precision nonlinear
susceptibility measurement. Next, we give the sensitivity of the nonlinear
phase shift.

\subsection{Homodyne detection}

Here, we consider the homodyne detection as our measuring method. The phase
sensitivity is described by the relation%
\begin{equation}
\Delta^{2}\phi=\frac{\langle\Delta^{2}\hat{O}\rangle}{\left\vert
\partial\langle\hat{O}\rangle/\partial\phi\right\vert ^{2}},
\end{equation}
where $\langle\Delta^{2}\hat{O}\rangle$ is the fluctuation of the observable
$\hat{O}$, and $\partial\langle\hat{O}\rangle/\partial\phi$ is the slope with
respect to the corresponding phase shift. The detected variable can be phase
quadrature or the photon number. In our scheme, the observable is phase
quadrature $\hat{Y}_{a_{3}}=-i(\hat{a}_{3}-\hat{a}_{3}^{\dagger})$. For
convenience, the following $\phi$ all represent nonlinear phase shifts, and we
omit the subscript $n$.

The slope of the quadrature $\hat{Y}$\ is given by
\begin{align}
|\partial\langle\hat{Y}_{a_{3}}\rangle/\partial\phi| &  =|\langle g_{2}%
[\hat{a}_{0}g_{1}Te^{i(\theta_{2}-\theta_{1}-\phi(2\hat{n}_{d_{1}}+1))}%
+(\hat{b}_{0}^{\dagger}G_{1}T\nonumber\\
&  +\hat{c}_{0}^{\dagger}\sqrt{TR})e^{i(\theta_{2}-\phi(2\hat{n}_{d_{1}}%
+1))}](2\hat{n}_{d_{1}}+1)\nonumber\\
&  +\text{H.c.}\rangle|.
\end{align}
For convenience, we analyze the phase sensitivity at $\phi=0$. When the two
input ports of the SU(1,1) interferometer have no injection, i.e., $\hat
{a}_{0}$ and $\hat{b}_{0}$ are in vacuum states, and the pump light is in a
coherent state $\left\vert \alpha\right\rangle $ with $\alpha=|\alpha
|e^{i\theta_{\alpha}}$ where $\alpha$ is a complex number and $\theta_{\alpha
}$ is the initial phase, then the slope\ of the quadrature $\hat{Y}$\ is
reduced to%
\begin{align}
|\partial\langle\hat{Y}_{a_{3}}\rangle/\partial\phi| &  =2g_{2}\sqrt
{TR}N_{\alpha}^{1/2}\left(  1+2RN_{\alpha}+4Tg_{1}^{2}\right)  \nonumber\\
&  \times|\cos(\theta_{2}-\theta_{\alpha})|,\label{eq7}%
\end{align}
where $N_{\alpha}=|\alpha|^{2}$. The increase in the slope of the output
signal is due to the amplification of the second nonlinear process and the
increase in phase-sensitive photons due to the strong pumping field when
$|\cos(\theta_{2}-\theta_{\alpha})|=1$. The corresponding fluctuation\ is
given by%
\begin{align}
\langle\Delta^{2}\hat{Y}_{a_{3}}\rangle &  =G_{2}^{2}G_{1}^{2}+g_{1}^{2}%
g_{2}^{2}+G_{2}^{2}g_{1}^{2}+G_{1}^{2}g_{2}^{2}\nonumber\\
&  +4G_{2}G_{1}g_{1}g_{2}\cos(\theta_{2}-\theta_{1}).\label{eq8}%
\end{align}

From Eq.~(\ref{eq7}) and Eq.~(\ref{eq8}), $R$ and $T$ are only related to the
slope and are independent of noise. The optimal $R/T$ is given by
\begin{align}
\frac{R}{T} &  =1/(2N_{\alpha}+1)[3N_{\alpha}-6g_{1}^{2}\nonumber\\
&  +\sqrt{9N_{\alpha}^{2}-28N_{\alpha}g_{1}^{2}+2N_{\alpha}+36g_{1}^{4}%
+4g_{1}^{2}+1}],
\end{align}
which is independent of $g_{2}$. When the intensity of the input coherent
state is strong enough, we obtain the optimal
\begin{equation}
\left(  \frac{R}{T}\right)  _{optimal}\approx3,
\end{equation}
which is because the nonlinear term $4R\sqrt{TR}N_{\alpha}^{3/2}$ in Eq.
(\ref{eq7}) dominates\ in phase sensitivity. Different forms of nonlinear
phase shifts produce different $R/T$ optimal ratios. If the nonlinear phase
shift is not considered, Eq.~(\ref{eq7}) can be rewritten as $|\partial
\langle\hat{Y}_{a_{3}}\rangle/\partial\phi|=2g_{2}\sqrt{TR}N_{\alpha}%
^{1/2}|\cos(\theta_{2}-\theta_{\alpha})|$. Under this condition, the optimum
value of $R/T$ is $1$, which is the commonly used ratio in the MZI for linear
phase-shift estimation.

Our scheme can also be thought of as inserting an MZI into one of the arms of
the SU(1,1) interferometer. To illustrate the quantum correlation enhancement
effects, we first consider a balanced case $G_{1}=G_{2}$, $\theta_{\alpha}=0$,
$\theta_{1}=0$ and $\theta_{2}=\pi$. The phase sensitivity of our scheme is%
\begin{equation}
\Delta\phi_{\text{balance}}=\frac{1}{g\left(  \mathcal{T}_{Lin}+\mathcal{T}%
_{Nonlin}+\mathcal{T}_{Nonlin\&Corr}\right)  },\label{eq9}%
\end{equation}
where $\mathcal{T}_{Lin}=2\sqrt{TR}N_{\alpha}^{1/2}$, $\mathcal{T}%
_{Nonlin}=4R\sqrt{TR}N_{\alpha}^{3/2}$, and $\mathcal{T}_{Nonlin\&Corr}%
=4T\sqrt{TR}N_{\alpha}^{1/2}N_{g}$. $N_{g}=2g^{2}$ is the spontaneous photon
number emitted from the NBS, which is related to parametric strength.
Additionally, the second NBS further enhances the phase\ sensitivity
by\ introducing an overall\ prefactor $g$ to the slope.

If we only consider the MZI in the dashed box, Eq. (\ref{eq9}) is simplified
to $\Delta\phi_{\text{MZI}}=1/(\mathcal{T}_{Lin}+\mathcal{T}_{Nonlin})$.
Furthermore, when the nonlinear phase shift is replaced by the linear phase
shift, $\Delta\phi_{\text{MZI}}$ is reduced to $1/\mathcal{T}_{Lin}$. Compared
to the usual MZI of linear phase shift, $\mathcal{T}_{Nonlin\&Corre}$ results from
the combined action of nonlinear phase shift and quantum correlation created
by the first NBS.

Next, we compare the optimal phase sensitivity of our scheme with the SQL. The
number of phase-sensing photons of the scheme is
\begin{equation}
N_{ps}=N_{g1}+N_{\alpha},
\end{equation}
where $N_{g1}=2g_{1}^{2}$.
For a conventional SU(1,1) interferometer with vacuum state input, $N_{ps}$ is $N_{g1}$. The MZI can tolerate a large
number of photons $N_{\alpha}$, and the output light does not affect the
quantum correlation when the small phase shift around $\phi=0$ is considered.
Therefore, the intrinsic problem of a small number of phase-sensing photons is
solved. The SQL for the nonlinear phase shift is $1/N_{ps}^{3/2}$ \cite{N2}. The phase
sensitivity as a function of $g_{2}/g_{1}$\ is shown in Fig. \ref{fig2} at
$\phi=0$. When $g_{1}$ and $N_{\alpha}$ are given, the value of $g_{2}/g_{1}$
is $2$ or greater, and the phase sensitivity reaches a stable
optimal value. The phase sensitivity obtained by the homodyne detection can
beat the SQL.

\subsection{Quantum Cram\'{e}r-Rao bound}

The quantum Fisher information (QFI) is the intrinsic information in the
quantum state and is not related to the actual measurement procedure. The QFI is
at least as great as the classic Fisher information for the optimal
observable, which gives an upper limit to the precision of quantum parameter
estimation. The QCRB according to the QFI is given by%
\begin{equation}
\Delta\phi_{\mathcal{F}}=\frac{1}{\sqrt{m\mathcal{F}}}, \label{QCRB}%
\end{equation}
where $m$ is the number of independent repeats of the experiment.

The QFI $\mathcal{F}$ is defined as $\mathcal{F}=\mathrm{Tr}[\rho(\phi
)L_{\phi}^{2}]$ \cite{Braunstein94,Braunstein96}, where the Hermitian operator
$L_{\phi}$, called the symmetric logarithmic derivative, is defined as the
solution of the equation $\partial_{\phi}\rho(\phi)=[\rho(\phi)L_{\phi
}+L_{\phi}\rho(\phi)]/2$. In terms of the complete basis $\{|k\rangle\}$ such
that $\rho(\phi)=\sum_{k}p_{k}|k\rangle\langle k|$ with $p_{k}\geq0$ and
$\sum_{k}p_{k}=1$, the QFI can be written as $\mathcal{F}=\sum_{k,k^{\prime}%
}\frac{2}{p_{k}+p_{k^{\prime}}}\left\vert \langle k|\partial_{\phi}\rho
(\phi)|k^{\prime}\rangle\right\vert ^{2}$
\cite{Braunstein94,Braunstein96,Toth}. Under the lossless condition, for a
pure state, the QFI is reduced to $\mathcal{F}=4(\langle\Psi_{\phi}^{\prime
}|\Psi_{\phi}^{\prime}\rangle-|\langle\Psi_{\phi}^{\prime}|\Psi_{\phi}%
\rangle|^{2})$, where the state after the phase shift is represented as
$|\Psi_{\phi}\rangle$ and $|\Psi_{\phi}^{\prime}\rangle=\partial|\Psi_{\phi
}\rangle/\partial\phi$.

Under the condition of $g_{1}=g$ and $\ \theta_{1}=0$, and considering the
nonlinear phase shift $\hat{d}_{2}=e^{i\phi(2\hat{n}_{d_{1}}+1)}\hat{d}_{1}$,
compared to the linear phase shift, the QFI $\mathcal{F}$ is given by%
\begin{equation}
\mathcal{F}=4\left[  \left\langle \hat{n}_{d_{1}}^{4}\right\rangle
-\left\langle \hat{n}_{d_{1}}^{2}\right\rangle ^{2}\right]  =N_{\alpha}%
^{3}s_{1}+N_{\alpha}^{2}s_{2}+N_{\alpha}s_{3}+s_{4},
\end{equation}
where%
\begin{align}
s_{1}  &  =16R^{4}+16R^{3}T(N_{g}+1),s_{2}=24R^{4}+R^{3}T(88N_{g}\nonumber\\
&  +48)+R^{2}T^{2}(52N_{g}^{2}+88N_{g}+24),\nonumber\\
s_{3}  &  =4R^{4}+R^{3}T(52N_{g}+12)+R^{2}T^{2}(96N_{g}^{2}+104N_{g}%
\nonumber\\
&  +12)+RT^{3}(40N_{g}^{3}+96N_{g}^{2}+52N_{g}+4),\nonumber\\
s_{4}  &  =T^{4}(5N_{g}^{4}+16N_{g}^{3}+13N_{g}^{2}+2N_{g})+T^{3}R(16N_{g}%
^{3}\nonumber\\
&  +26N_{g}^{2}+6N_{g})+T^{2}R^{2}(13N_{g}^{2}+6N_{g})+2TR^{3}N_{g}.
\end{align}
If only considering the linear phase shift, the above QFI $\mathcal{F}$ can be
reduced to%
\begin{align}
\mathcal{F}  &  =4\left\langle \Delta^{2}\hat{n}_{d_{1}}\right\rangle
=N_{\alpha}[4R^{2}+4RT(N_{g}+1)]\nonumber\\
&  +N_{g}[T^{2}N_{g}+2TR]+2T^{2}N_{g}.
\end{align}
Under the condition of $N_{\alpha}=0$ and $T=R=1/2$, $\mathcal{F}$ is given by
$\mathcal{F}=\frac{1}{4}[N_{g}(N_{g}+2)+2N_{g}]$. Compared to the form
$N_{g}(N_{g}+2)$ in an SU(1,1) interferometer \cite{Li16}, the prefactor $1/4$
results from the introduction of a BS in one arm of the SU(1,1) interferometer,
and the extra term $2N_{g}$ is due to the vacuum fluctuation despite
$N_{\alpha}=0$.

Next, we compare the QCRB with the SQL, and the phase sensitivity obtained by
the homodyne detection. As shown in Fig.~\ref{fig2}, the optimal phase
sensitivity obtained by homodyne detection can beat the SQL and can approach
the QCRB.

\section{Losses}

In the presence of realistic imperfections, the ultimate precision limit in
noisy quantum-enhanced metrology was also studied. In this section, we
investigate the effects of losses on phase sensitivities.

\subsection{Internal and external losses of MZI}

\begin{figure}[ptb]
\includegraphics[scale=0.45,angle=0]{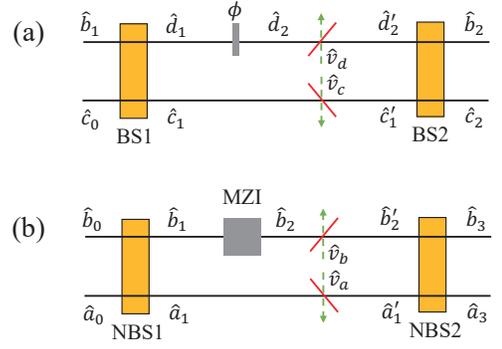}
\caption{ A lossy
interferometer model, the losses in the interferometer are modeled by adding
fictitious beam splitters. (a) Internal losses. (b) External losses.}%
\label{fig3}%
\end{figure}

Losses can be modeled by adding fictitious beam splitters, as shown in
Fig.~\ref{fig3}. Considering both arms of the MZI have different internal
transmission rates $\eta_{c}$ and $\eta_{d}$, and outside transmission rates
$\eta_{a}$ and $\eta_{b}$, the mode transforms of the fields are given by
\begin{align}
\hat{c}_{1}^{\prime}  &  =\sqrt{\eta_{c}}\hat{c}_{1}+\sqrt{1-\eta_{c}}\hat
{v}_{c},\text{ }\hat{d}_{2}^{\prime}=\sqrt{\eta_{d}}\hat{d}_{2}+\sqrt
{1-\eta_{d}}\hat{v}_{d},\\
\hat{a}_{1}^{\prime}  &  =\sqrt{\eta_{a}}\hat{a}_{1}+\sqrt{1-\eta_{a}}\hat
{v}_{a},\text{ }\hat{b}_{2}^{\prime}=\sqrt{\eta_{b}}\hat{b}_{2}+\sqrt
{1-\eta_{b}}\hat{v}_{b},
\end{align}
where $\hat{v}_{a}$, $\hat{v}_{b}$, $\hat{v}_{c}$, and $\hat{v}_{d}$ represent
the vacuum. Considering the losses, the input-output relation of $\hat{a}_{3}$
is%
\begin{equation}
\hat{a}_{3_{l}}=\hat{a}_{0}\mathcal{A}_{l}+\hat{b}_{0}^{\dagger}%
\mathcal{B}_{l}+\hat{c}_{0}^{\dagger}\mathcal{C}_{l}+\hat{v}_{a}%
\mathcal{V}_{1}+\hat{v}_{b}^{\dagger}\mathcal{V}_{2}+\hat{v}_{c}^{\dagger
}\mathcal{V}_{3}+\hat{v}_{d}^{\dagger}\mathcal{V}_{4},
\end{equation}
with
\begin{align}
\mathcal{A}_{l}  &  =\sqrt{\eta_{a}}G_{2}G_{1}+\sqrt{\eta_{b}}g_{2}%
g_{1}e^{i\left(  \theta_{2}-\theta_{1}\right)  }(\sqrt{\eta_{d}}%
Te^{-i\phi(2\hat{n}_{d_{1}}+1)}\nonumber\\
&  +\sqrt{\eta_{c}}R),\nonumber\\
\mathcal{B}_{l}  &  =\sqrt{\eta_{a}}G_{2}g_{1}e^{i\theta_{1}}+\sqrt{\eta_{b}%
}G_{1}g_{2}e^{i\theta_{2}}(\sqrt{\eta_{d}}Te^{-i\phi(2\hat{n}_{d_{1}}%
+1)}\nonumber\\
&  +\sqrt{\eta_{c}}R),\nonumber\\
\mathcal{C}_{l}  &  =\sqrt{\eta_{b}}g_{2}e^{i\theta_{2}}\sqrt{TR}(\sqrt
{\eta_{d}}e^{-i\phi(2\hat{n}_{d_{1}}+1)}-\sqrt{\eta_{c}}),\nonumber\\
\mathcal{V}_{1}  &  =G_{2}\sqrt{1-\eta_{a}},\text{ }\mathcal{V}_{2}%
=g_{2}e^{i\theta_{2}}\sqrt{1-\eta_{b}},\nonumber\\
\mathcal{V}_{3}  &  =-g_{2}e^{i\theta_{2}}\sqrt{R}\sqrt{\eta_{b}(1-\eta_{c}%
)},\text{ }\nonumber\\
\mathcal{V}_{4}  &  =g_{2}e^{i\theta_{2}}\sqrt{T}\sqrt{\eta_{b}(1-\eta_{d})},
\end{align}
where subscript $l$ indicates the loss. Similar to the lossless case, we
analyze the phase sensitivity at $\phi=0$. At this phase point, the slope is%
\begin{align}
|\partial\langle\hat{Y}_{a_{3_{l}}}\rangle/\partial\phi|  &  =2g_{2}\sqrt
{\eta_{b}\eta_{d}TR}N_{\alpha}^{1/2}\nonumber\\
&  \times\left(  1+2RN_{\alpha}+2TN_{g}\right)  |\cos(\theta_{2}%
-\theta_{\alpha})|,
\end{align}
and the fluctuation of the quadrature $\hat{Y}_{a_{3_{l}}}$ is%
\begin{align}
\langle\Delta^{2}\hat{Y}_{a_{3_{l}}}\rangle &  =\eta_{a}G_{2}^{2}G_{1}%
^{2}+\eta_{b}g_{2}^{2}g_{1}^{2}\left(  \sqrt{\eta_{d}}T+\sqrt{\eta_{c}%
}R\right)  ^{2}\nonumber\\
&  +\eta_{a}G_{2}^{2}g_{1}^{2}+\eta_{b}g_{2}^{2}G_{1}^{2}\left(  \sqrt
{\eta_{d}}T+\sqrt{\eta_{c}}R\right)  ^{2}\nonumber\\
&  +\eta_{b}g_{2}^{2}TR\left(  \sqrt{\eta_{d}}-\sqrt{\eta_{c}}\right)
^{2}\nonumber\\
&  +\left[  \left(  1-\eta_{a}\right)  G_{2}^{2}\right]  +\left[  \left(
1-\eta_{b}\right)  g_{2}^{2}\right] \nonumber\\
&  +\left[  \eta_{b}\left(  1-\eta_{c}\right)  g_{2}^{2}R\right]  +\left[
\eta_{b}\left(  1-\eta_{d}\right)  g_{2}^{2}T\right] \nonumber\\
&  +4\sqrt{\eta_{a}\eta_{b}}G_{2}G_{1}g_{1}g_{2}\left(  \sqrt{\eta_{d}}%
T+\sqrt{\eta_{c}}R\right)  \cos(\theta_{2}-\theta_{1}).
\end{align}
From the above two equations, the condition for obtaining the optimal phase
sensitivity is $\theta_{\alpha}=0$, $\theta_{1}=0$, and $\theta_{2}=\pi$. We
can study the effects of internal losses by setting $\eta_{a}=\eta_{b}=1$, or
study the effects of the external losses by setting $\eta_{c}=\eta_{d}=1$.

\begin{figure}[ptb]
\includegraphics[scale=0.4,angle=0]{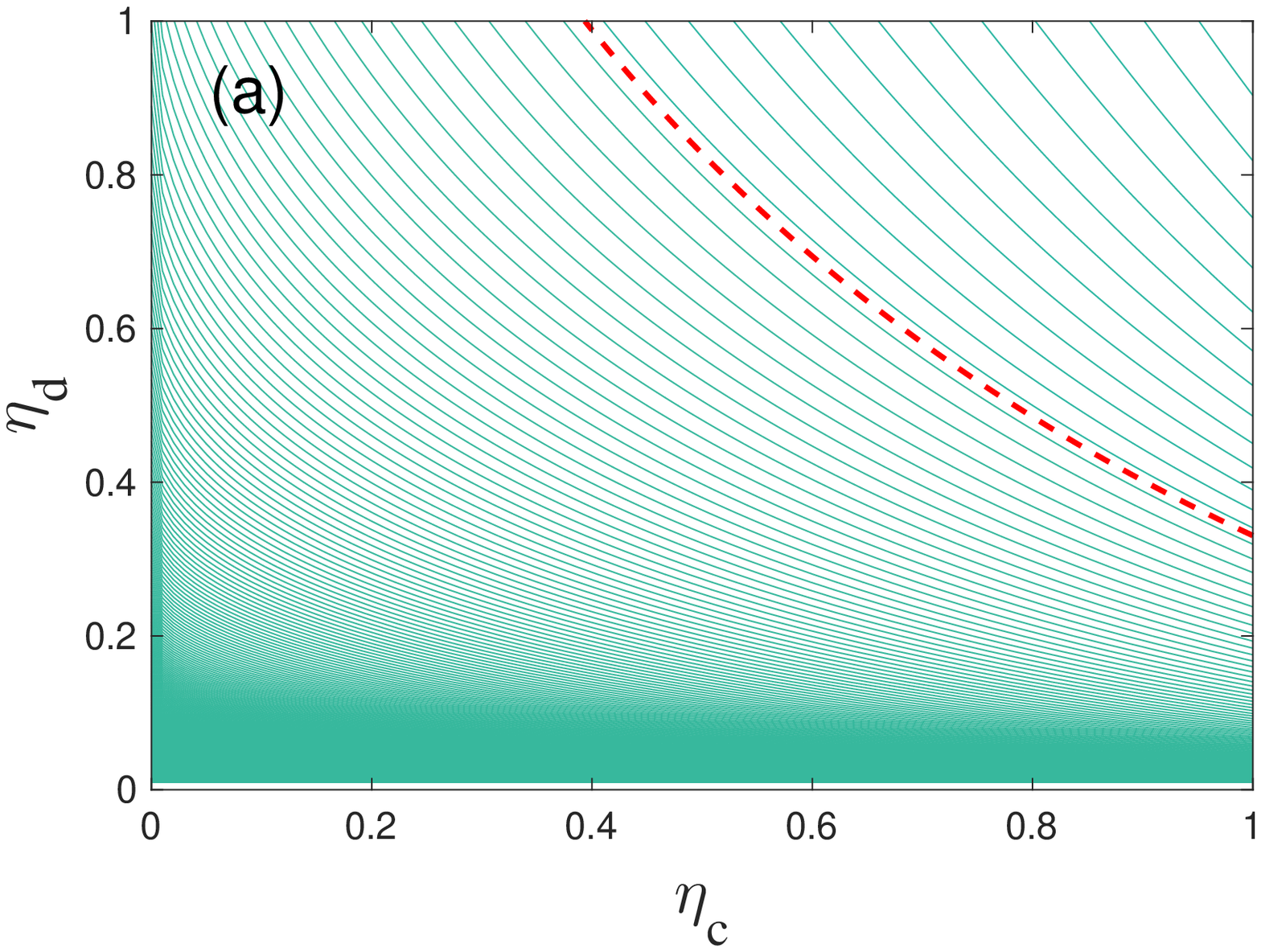}
\includegraphics[scale=0.4,angle=0]{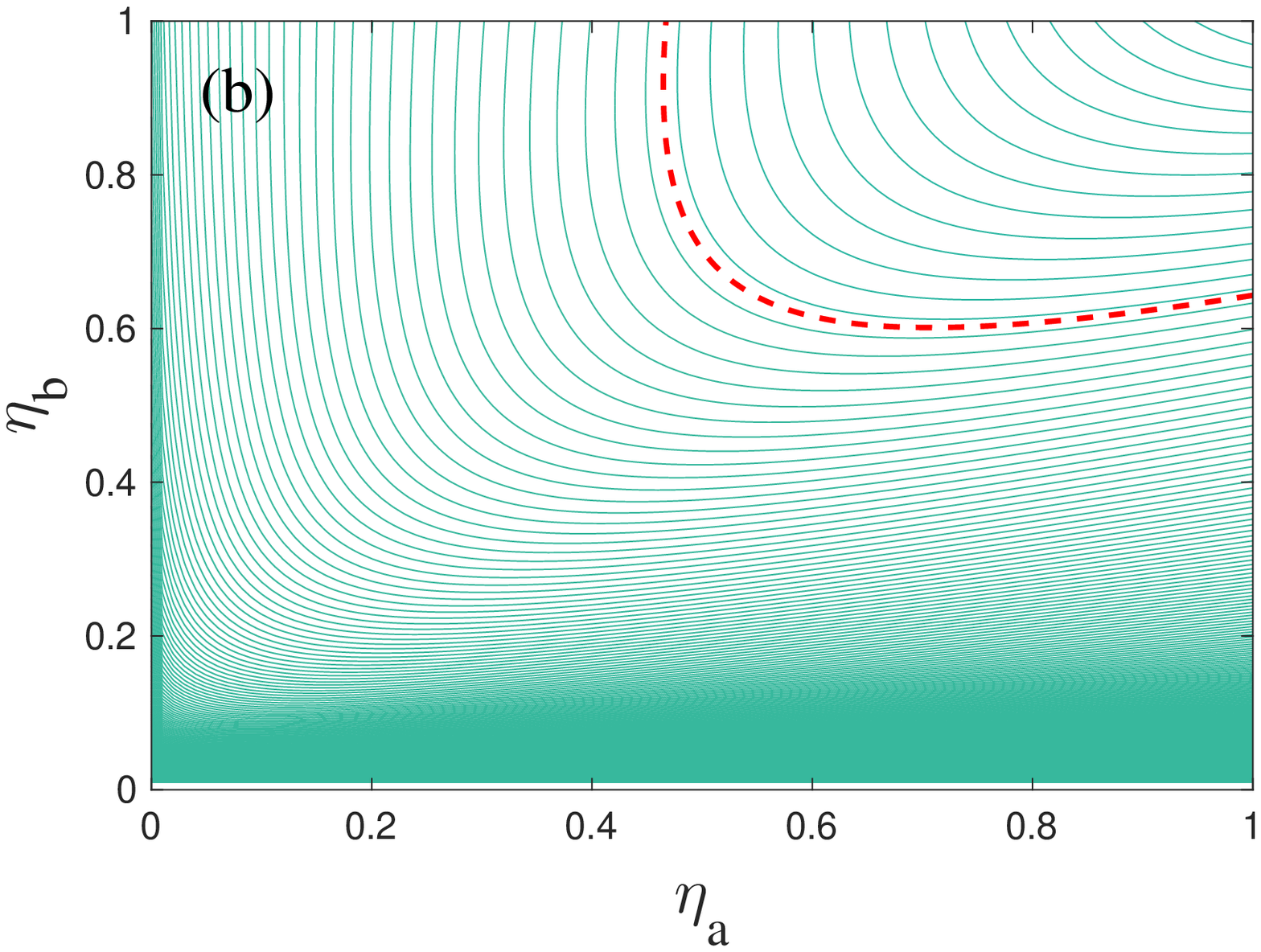} \caption{ Phase sensitivity
vs photon loss coefficient $\eta$, where $|\alpha|=10$, $g_{1}=2$, and
$R/T=3$. (a) Internal losses. (b)
External losses. The dashed lines denote $1/N_{ps}^{3/2}$. The phase sensitivities in the upper right corner and within
the dashed line area can overcome $1/N_{ps}^{3/2}$. }%
\label{fig4}%
\end{figure}

For convenience, we first consider $g_{2}=2g_{1}$. The phase sensitivity as a
function of photon losses in both arms $\eta_{c}$ and $\eta_{d}$ when
$\eta_{a}=\eta_{b}=1$ is shown in Fig.~\ref{fig4}(a), where the dashed line
denotes $1/N_{ps}^{3/2}$. The phase sensitivities in the upper right corner
and within the dashed line area can overcome $1/N_{ps}^{3/2}$. Since the
strong pump field injection and the unbalanced BS make the intensity of the
upper arm of the MZI larger than that of the lower arm, it is more tolerant
with loss of light field $\eta_{d}$. It is shown that the interferometer can
tolerate approximately $70\%$ of the photon losses when $\eta_{c}=1$.
Similarly, considering external photon losses, the phase sensitivity as a
function of photon losses in both arms $\eta_{a}$ and $\eta_{b}$ when
$\eta_{c}=\eta_{d}=1$\ is shown in Fig.~\ref{fig4}(b). It is demonstrated that
the phase sensitivity can still overcome $1/N_{ps}^{3/2}$ with $40\%$ of the
photon losses. Large photon loss affects quantum correlation, which reduces
the sensitivity of measurement.

\subsection{Detection losses}

Finally, we study the detection loss, and the field undergoing the detection
losses is $\hat{a}_{3}^{^{\prime}}=\sqrt{\eta}\hat{a}_{3}+\sqrt{1-\eta}\hat
{v}$, where $\hat{v}$ is the vacuum. The corresponding slope and variance are%
\begin{equation}
|\partial\langle\hat{Y}^{^{\prime}}\rangle/\partial\phi_{n}|=2g_{2}%
\sqrt{TR\eta}N_{\alpha}^{1/2}\left(  1+2RN_{\alpha}+4Tg_{1}^{2}\right)  ,
\end{equation}
and
\begin{equation}
\langle\Delta^{2}\hat{Y}^{^{\prime}}\rangle=\eta\lbrack(G_{2}G_{1}-g_{2}%
g_{1})^{2}+(G_{2}g_{1}-G_{1}g_{2})^{2}]+1-\eta,
\end{equation}
respectively. For a balanced configuration ($G_{1}=G_{2}$, $\theta_{\alpha}%
=0$, $\theta_{1}=0$, and $\theta_{2}=\pi$), this type of loss only introduces a
prefactor $1/\sqrt{\eta}$ to the sensitivity, i.e., $\Delta\phi^{\prime
}=(\Delta\phi)_{\text{balance}}/\sqrt{\eta}$, which is the same as the result
by intensity detection \cite{Marino12}.

\section{Conclusion}

In conclusion, we have proposed a scheme to enhance the nonlinear phase
estimation of an MZI with a coherent state in one input port and one mode of a
two-mode squeezed-vacuum state in the other input port and with active
correlation readout. The phase sensitivity is improved compared to the
traditional MZI for the same input phase sensing field because of active
correlation output readout. Due to introduction of nonlinear phase estimation, the optimal $R/T$ ratio is about $3$ instead of $1$ usually used under the condition of strong coherent-state input. The phase sensitivity can beat the SQL
and approach the QCRB using the method of homodyne detection under the
condition of the optimal $R/T$ ratio. The internal and external losses of the
optical field degraded the measurement precision, and we have given their
critical values where the phase sensitivity is below the SQL in the presence
of photon losses. The detection loss only introduces a prefactor $1/\sqrt
{\eta}$ to the sensitivity for the balanced configuration. Due to the relation
between the nonlinear phase shift and the susceptibility $\chi^{(3)}$ of the
Kerr medium, our scheme can also provide the estimation of nonlinear
susceptibility $\chi^{(3)}$.

\section{ACKNOWLEDGMENTS}

This work is supported by National Natural Science Foundation of China
Grants No. 11974111, No. 11874152, No. 91536114, No. 11574086,
NO. 11974116, No. 11654005, and No. 11474095; Development Program of China Grant No.
2016YFA0302001; Natural Science Foundation of Shanghai Grant No. 17ZR1442800;
Shanghai Rising-Star Program Grant No. 16QA1401600; and the Fundamental Research
Funds for the Central Universities.

\end{document}